\documentclass{article}[12pt]
\usepackage[T1,T2A]{fontenc}
\usepackage[utf8]{inputenc}
\usepackage{amsmath,amssymb}
\usepackage{graphicx}%
\usepackage{xcolor}
\usepackage{verbatim}
\usepackage{multirow}

\newcommand{\FIG}[3]{%
\begin{center}
\parbox{#2cm}{%
\refstepcounter{figure}\includegraphics[width=#2cm]{#1} \noindent Fig. \thefigure:\quad
#3}\end{center}}

\newcommand{\RisTextReg}[5]{%
\begin{center}
\begin{tabular}{lr}
\parbox{#2cm}{\includegraphics[width=#2cm]{#1}\\
\refstepcounter{figure} {\bf Fig. \thefigure.} #3} &
\parbox{#4cm}{#5}\\
\end{tabular}\end{center}
\vspace{8pt}%
}
\newcounter{strochka}

\newcounter{spisok}
\begin{document}

\begin{center}
{\bf \Large Yurii Ignat'ev\footnote{Institute of Physics, Kazan Federal University, 420008 Russia, Kazan, Kremlevskaya st. 18; email: yurii.ignatev.1947@yandex.ru} }\\[12pt]
{\bf \LARGE From Geometry to Physics -- AZ: touches to the portrait} \\[12pt]
\end{center}

\abstract{The process launched by Lobachevsky. The movement of the Kazan school of geometry towards physics. Personal memories of Alexei Zinovievich Petrov, the great Kazan geometer and theoretical physicist, who became the Author's Guiding Star. \\

{\bf KeyWords}: Alexey Zinovievich Petrov, geometry, general theory of relativity, Kazan University, Department of Relativity and Gravitation Theory, methods of teaching exact sciences.
}

\section*{Big things are seen from a distance}
The proverbs like ``Big things are seen from a distance''\footnote{Russian proverb} and ``You can't see the forest for the trees''\footnote{Russian proverb} should be close to the heart of a theoretical physicist. Indeed, particles that are close to each other experience strong interparticle interactions, they are dressed in their fur coat, so it is difficult to single out their own characteristics. In order to see them, the particles need to be moved further away from each other, then the concepts of their own mass, charge, etc. will appear. Also, being inside the Galaxy, in its field of influence, it is difficult to understand its structure, and only by examining similar, in our opinion, distant objects, whose nuclei are not hidden from our sight by dust nebulae, can we form an idea of our Galaxy and understand that the silvery-diamond stripe passing on the southern slope of the night sky is the center of our Galaxy.

\RisTextReg{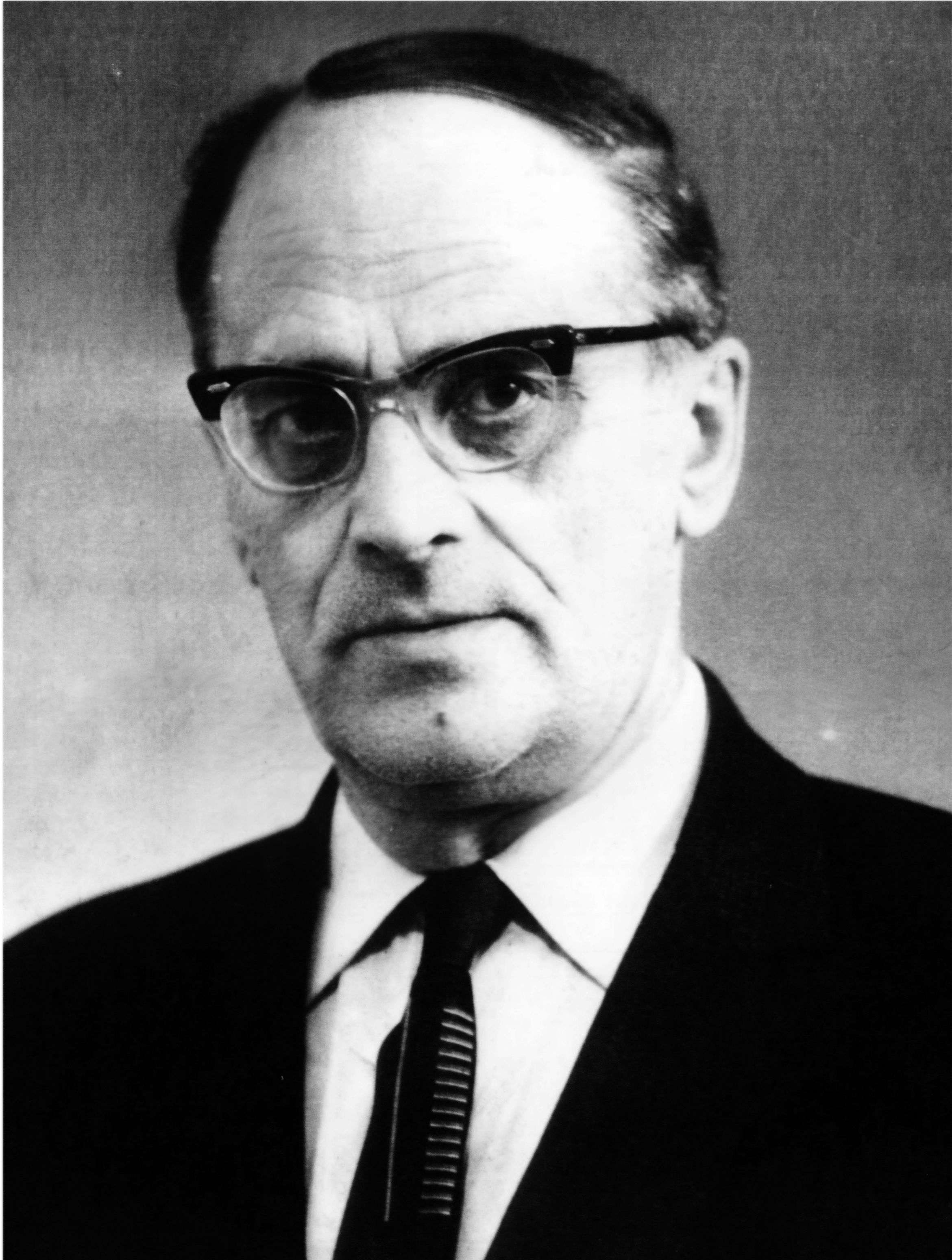}{7.5}{Alexey Zinovievich Petrov. 1910--1973}{8}{
So, remembering Alexey Zinovievich Petrov 50 years after the break of his world line in this world, you begin to involuntarily realize all the greatness of this man and understand that this silver-diamond path on the horizon of our scientific life is what he left behind - the center of our Galaxy. Personally, I have another reason for farsightedness - I caught Alexey Zinovievich in the prime of his glory, being at that time an ordinary student of the physics department of Kazan University, and for seven years I was in the area of the powerful force field of his personality. For me in those years it was completely ordinary that an extraordinary person, a world-class scientist, communicated with me, took part in my fate. The great is seen from a distance. When his world line broke\footnote{Here I involuntarily turn to the vivid comparison of George Gamov \cite{Gamov}, since it is difficult to think of a better analogy.}, I felt very lonely and I began to understand all the greatness of the past moment, which gave me the light of a bright star.

And now I want to add some features to the well-known academic portrait of AZ\footnote{that's what his colleagues and students called him}, supplementing it with particles of my personal experience, which, perhaps, will enliven this portrait and make it richer. Living life, you gradually begin to understand that such details are precisely what are important for the formation of personality, and, it would seem, important details of its environment have very little influence on the trajectory of its flight. "It ain't the roads we take; it's what inside of us that makes us turn out the way we do."\ (O$'$ Henry).
}

Nevertheless, I will try to reproduce some colorful details of the environment of that time, especially since these details help to recreate the stage on which Alexey Zinovievich Petrov performed in the 60-70s of the 20th century.

\section{First meeting with Petrov and choice of the Physics Department: 1963-1965}
I met Alexey Zinovievich Petrov in absentia in the fall of 1963, when I was in the 10th grade of an 11-year school and visited the library on Lenin Street (Aquarium), where I read popular science literature on astronomy and physics. There I came across a brochure by A.Z. Petrov, Space - Time and Matter: An Elementary Outline of the Modern Theory of Relativity \cite{AZ_brosh}. The brochure very clearly and briefly outlined the basic concepts of the theory of relativity. I was surprised to learn that the author of this brochure worked at Kazan University. The brochure made a strong impression on me, and as a result I went into the lobby of the main building of Kazan University, where I bought this book at a book kiosk. Here I learned that the KSU Physics Department has the only department of relativity and gravitation theory in the USSR, headed by A.Z. Petrov. Since then, all my thoughts turned to the Physics Department of Kazan University. In the summer of 1965, when I was preparing for the entrance exams to the Physics Department of KSU, an article was published in the newspaper "Soviet Tatarstan" with a title approximately like this "Scientists from Kazan University have discovered gravitational waves". This article reported that a group of scientists from the Department of Relativity and Gravitation Theory, led by Professor A.Z. Petrov, had discovered gravitational waves on a gravitational detector. This article immediately made me, casting aside all hesitation, enroll in the Physics Department, and then in the Department of Relativity and Gravitation Theory. I remember how my heart was beating with joyful excitement - after all, I was right there, very close to people doing real science! As it turned out later, this was an erroneous and very, very premature publication, but the deed was done, and it was a good deed that determined my future destiny.

In August 1965, I passed the entrance exams to the Physics Department (24.5 points out of 25 possible) and was enrolled in the first year of the Physics Department. The competition for the Physics Department that year was very high, also due to the fact that secondary schools simultaneously graduated two classes - with a ten-year and eleven-year education. In addition, applicants in the 60s were very motivated to study and were strong. The passing score at the Physics Department was 23.5, i.e., it was necessary to get at least 3 A's, one 4.5 and one B. It should be noted that in those years, grades were given at the entrance exams to the Physics Department with an accuracy of 0.5 points. I got a 4.5 on the written math exam, having corrected the situation on the oral exam, when after listening to my answer the examiner added 0.5 points to my written work. I waited for the classes to start with joyful excitement. But at the very first lecture a serious teacher came in and announced that we were all going for a month ``to pick potatoes'' and explained to us the procedure for preparing for the trip.\footnote{In those years and up until 1991, there was a "pleasant"\ tradition in universities, when students spent a month doing work training, mainly on collective farms.}
\section{Petrov's Lectures and Methodology: 1966-1967}
After a month of picking beets in the Salmanovsky district of Tatarstan under almost continuous rain and the process of "group coordination"\ in the clay creeping underfoot and spending the night in huts on the floor in groups of 8 people, we turned from inexperienced youths who did not know the taste of wine into seasoned, as it seemed to us then, students. We returned to the university as a united "653rd armored tank"\ group (the group acquired this name in the student environment for its cohesion and high activity, in particular, opposition to the functionaries of the university Comsomol committee). In addition to standard course lectures, we also had overview lectures given by leading specialists of the physics department, talking about the scientific directions they led. Alexey Zinovievich repeatedly gave such lectures, and he gave them with obvious pleasure. At his lecture, the 1st physics auditorium (the largest of the lecture halls) was packed to capacity.

AZ, as he was respectfully called by both students and teachers, was a brilliant lecturer - he lectured leisurely, clearly, with articulation and his invariable ``p\textbf{\'o}etomu''  (that's why)\footnote{He seemed to be deliberately emphasizing the wrong stress, which all his students imitated.}. During the lecture, he walked sedately along the huge board, sometimes writing down the most important things on it, sucking on a cigarette and then, putting it on the chalk shelf, lighting another one. The lectures were very interesting, AZ often told about his meetings with famous scientists, including telling anecdotal stories, and he always behaved sedately, with dignity. I will tell you one such.

\subsection{What is the time?}
During a break between sessions of the IV Gravitational Conference in London in 1965, AZ was walking along the Thames embankment with V.A. Fock. Fearing to be late for the conference session and not having a watch with him, Fock asked a random passerby: ``What is the time?'' (instead of ``What time is it'' - Fock did not speak English very well). In response, the passerby thoughtfully replied: ``You know, I think about it all the time too.'' It turned out that he was a participant of the same gravitational conference, a well-known theoretical physicist.

\subsection{Girls and dancing}
AZ was always neat, elegant and even imposing, the female students looked at him with loving eyes. AZ himself did not ignore beautiful girls either. As Professor Yu. S. Vladimirov, who worked as Petrov's secretary in the Soviet Gravitational Commission, notes, "Nothing human was alien to him" (\cite{Vladimir_KazCas}, \cite{Vladimir_STFI}). AZ, on the one hand, was democratic and accessible, on the other, he always maintained a distance with the help of a certain shining halo of a celestial being, clearly visible to others. The students also respected him very much, they were impressed by these qualities of Petrov. More than once I noticed him at the student balls of the Physics Department, which were held in the gym on the second floor of the main building of the university, where the administrative offices are now located. I, like many unfledged students, did not know how to dance then, so we stood in a crowd against the wall, looking at the dancing couples, just like in the song "The girls are standing, standing to the side, fiddling with their handkerchiefs" (only the other way around). Alexey Zinovievich, having chosen a worthy partner for himself, very gallantly invited her to a waltz and waltzed very gracefully to the envy of his girlfriends. Well, the guys and I looked at this spectacle with admiration. I could not even imagine then that before this was a village boy with a very difficult fate. From the memoirs of A.Z. Petrov's classmate, Kazan geometer associate professor V.G., - "Once, at an evening, seeing that I was not dancing, but timidly pressing myself against the wall, he said: "We must dance. Sometimes it is more important than mathematics." \cite{Amin_2}, \cite{Amin}.

\subsection{Methodological conference of the Physics Department on issues of student education}
At that time, the university enjoyed university freedoms that are scary to think about now. All positions were elected, department heads and teachers had the right to choose their own curriculum and teaching methods, an assistant could allow himself to engage in an unpleasant discussion with the Rector at a general meeting of the faculty. Jumping ahead in time 15 years, I will give a personal example of such university freedom. At that time, I was already a teacher in the Department of Relativity and Gravity Theory and a candidate of sciences. Once we were warned that a commission was coming from the USSR Ministry of Higher Education to study the state of affairs at Kazan University, and asked to be more attentive. During my lecture on mathematical analysis, the door to the lecture hall opened, and I saw on its threshold a rather impressive group of respectable people headed by the Rector, Professor Nuzhin. Nuzhin very politely explained to me that they were experts from the Ministry of Higher Education Institutions from Moscow and they would like to attend my lecture. At that time, there was a strict rule - no outsiders were allowed to attend classes without the lecturer's permission. I also politely explained to the Rector that I had 200 students (8 academic groups) in the auditorium and the presence of such a large group of people at the lecture would ruin the creative atmosphere of the lecture and distract the students. Nuzhin, smiling sheepishly, addressed the commission - "You see how strict our teachers are, let's not break the rules". After that, the entire group of experts, animatedly discussing the precedent, moved on. It should be noted that no "organizational conclusions"\ were made with respect to me by the university administration, although this story became widely known and I was later, at convenient times, jokingly, reminded of it.

The teachers, not being constrained by strict curricula (the entire subject program was written out by hand on one sheet at the beginning of the school year), gave lectures at different levels - among them were favorite lecturers (and these were, first of all, scientists), who improvised a lot, making their lectures very interesting, filled with original proofs and methods of solving problems, and there were also methodologists who read their lectures boringly from textbooks. In those days, students had the right to choose from the lecture courses with the same content those that they liked more and even take exams with an alternative lecturer. At the Physics Department, interesting and vivid lectures were given by professors K.P. Sitnikov, A.Z. Petrov, S.A. Altshuller, associate professors B.I. Kochelaev, Sh.Sh. Bashkirov, Moskvin, Kaigorodov and Ovchinnikov. The rest of the lectures were boring and completely carbon copies of standard textbooks. Therefore, the strongest students either listened to lectures in other lecture streams or independently studied the subject from books. This was especially true for the four-semester course of general physics and mathematical subjects. We often ran to parallel lectures on mathematical subjects at the Faculty of Mechanics and Mathematics to the famous scientists A.P. Norden, V.V. Morozov and A.V. Yablokov, since the mathematics teachers at the Physics Department were then, for the most part, very mediocre, as were the teachers of the general physics course, with the exception of K.P. Sitnikov.

The methodologists, realizing that they could not withstand the competition with leading lecturers and were losing strong students both in lectures and in placement, decided to fight back using administrative resources. They planned to deprive both students and teachers of university freedoms. Under the pretext of the need to strengthen discipline and methodological support for lecture courses, they decided to create, on the basis of the methods commission, a body to control both the students' studies and their attendance, and the methods of the teachers. The main blow of the methods commission was aimed precisely at the brightest scientists of the Physics and Mechanics and Mathematics Departments: Norden, Petrov, Altshuller and Yablokov. It must be said that throughout my entire teaching career I have repeatedly observed how all sorts of educational and methods commissions were created, usually highly ideological, to solve problems in favor of one group of people (usually the mediocre part of the team), pursuing the goal of establishing control over another part of the people with subsequent subordination to themselves by creating certain rules of the game.

Be that as it may, somewhere in May 1967, the so-called "methodological conference"\ of the Physics Department took place in the 3rd Physics Auditorium of the Main Building. The students, having learned about the upcoming event and its possible negative consequences for their future lives, became worried and went to the Dean of the Physics Department, Professor M.M. Zaripov, who had good contact with the students. The Dean made a decision to elect 2 deputies from each academic group with the status of observers to the methodological conference. Group meetings and elections were quickly organized. Ravil Nigmatullin and Yuri Ignat$'$ev were elected from the 653rd academic group of the Physics Department (2nd year). There was no room to swing a cat at the conference - student representatives sat in the gallery (without the right to vote, but not without the right to think). The first report was made by the Methodological Commission by its chairman, Associate Professor I.S. Fishman. The proposals formulated in the report caused a sharp negative reaction from the leading specialists of the faculty. World-famous scientists - professors A.P. Norden, A.Z. Petrov, S.A. Altshuller and others - sharply criticized the provisions of the report. In particular, Aleksey Zinovievich said that it is impossible to drive a student like a horse in a circus arena - this way you can only train an obedient performer, and not a critically thinking scientist. A heated discussion and verbal skirmish arose. After the conference, bitterly imagining the possible consequences of this methodological activity, I spent the night at home composing an Appeal to teachers and students and in the morning I hung it on the wall in the corridor of the Physics Department. Now we will give the floor to the chairman of the aforementioned methodological commission and provide him with comments in those places where this "word"\ is not entirely adequate.

\subsection{"Dazibao"\ (from the book \cite{Fishman})}
\textbf{"}At the university, as an associate professor and then a professor of the department, I devoted much time to social work. Thus, for 15 years I headed the methodological commission of the physics department. On my initiative, methodological conferences on various issues of pedagogical activity were convened annually, usually at the end of each academic year. The first conference was devoted to the topic "On the mastery of lecturing."\ It aroused the keenest interest, especially among the professors. The conference materials were published under my editorship and with my preface. Professors Altshuller, Norden, Petrov, Yablokov and others came to the conference. Here, there was one incident.

The day after the conference, I came to the university early in the morning and saw a "dazibao"\ on the wall in the corridor on the first floor of the physics department (at that time, this Chinese word was fashionable).\footnote{Fashionable is, to put it mildly, not quite the right word. It must be understood that at that time, a "cultural revolution" was being harshly carried out in China. Chinese Komsomol fanatics, the "Red Guards", smashed and destroyed cultural works and pasted pogrom leaflets "dazibao" everywhere calling for reprisals against figures of science and culture. In the Far East, they staged provocations, trying to swim across the Amur in huge crowds on junks and seize part of the USSR. "Red Guards roam and wander near the city of Beijing, and Red Guards search and rummage for old paintings. And it's not that the Red Guards like statues and paintings, but instead of statues there will be urns of the cultural revolution... Vladimir Vysotsky. " Thus, by calling my wall newspaper "dazibao", the respected professor thereby called me a Red Guard, i.e., an illiterate fanatic calling for reprisals against figures of science. An elegant logical substitution! A political accusation, dangerous in 1967, the year of the 50th anniversary of the October Revolution. Moreover, at that time the Author of the "Memoirs..." was, it seems, also the party organizer of the physics department.}

Here is what the anonymous author wrote: To professors, students!!! Yesterday, at a conference devoted to the problems of education, a heated debate broke out between the supporters of two programs for transforming teaching methods: the Fishman program and the Norden-Petrov-Altshuller program. The essence of Prof. Fishman's project consists of a minor reorganization of the university education system. The main idea of ??the project is to create a more powerful apparatus for forcing lazy students to be interested in science...

Here I should explain that in his speech at the conference, Prof. Norden voiced the thesis that students should be taught like kittens to swim, i.e. thrown into the water: if they float, good, if not, God bless them. In Prof. Altshuller's speech, it was said that it is not necessary to give good lectures - a good student will figure it out himself from books. As for me, taking into account the poor attendance of students at lectures, I oriented the teachers not only towards individual "smart guys", but also towards all students, since we had already accepted them to the university.

The author writes: "Yes, Comrade Fishman, the university should teach smart people, not the masses!"\
I thought that it was impossible to rely on the impulsive statements of venerable professors that the quality of lectures is not important, that painstaking work of the entire teaching staff is not required for education.
A number of the anonymous author's positions were indisputable. It is correct that "the university should graduate not technical engineers, but high-quality scientists." But no one objected to this.

The anonymous author ends his treatise with the words: "Comrade scientists, we are proud of you and believe in you. And you - students who, by the evil will of fate, accidentally found yourself within these walls, be conscious - after all, this is a very important matter."\ And the signature: "Student."\ At the end it is added: "A request to the bureaucrats not to disrupt!!!"\

"Dazibao"\ was glued together from 6 notebook sheets about 80 cm long. And yet "anonymous"\ did not remain so for long. I noticed in the evening that despite the "sold out"\ and overcrowding of the 3rd physics auditorium with teachers of the physics department, several students got in there. Here the respected professor, the author of "Memoirs", is mistaken. On the eve of the conference, considering its fateful character that was being prepared and did not promise anything good in the future, all academic groups were asked to choose two representatives for the conference as observers. From our friendly 653rd group, Ravil Nigmatullin (now a venerable theoretical physicist) and I were chosen. The student representatives were present at the conference on completely legal grounds. My suspicion fell on my students Leonid Nefedyev, now a prominent scientist, professor at the Kazan Pedagogical University, and Ignat$'$ev, now a professor at the same university. \footnote{Here the respected professor is wrong twice again. Before I, then still an associate professor, took the exam in optics, which was a semester part of the general physics course, he invited me and Ravil Nigmatullin to the optics department. Nefedyev was nowhere near there. In addition, by the time he wrote the "Memoirs", Ignatyev had already been the head of the department at the pedagogical institute for 18 years and an Honored Scientist of the Republic of Tatarstan, unlike Nefedyev.}. I took them "on the gun", said that the "dazibao"\ was the work of their hands or their comrades. Ignat$'$ev immediately admitted that he wrote it.

In my private conversation with Ignatiev, which took place soon after, we clarified our positions in detail and parted as good friends.\textbf{"}

\section{At the Department of Relativity and Gravitation: 1967 -- 1970}
\subsection{Distribution}
At the end of the 2nd year, the Physics Department distributed students among departments and corresponding specializations. The departments had systems of mandatory special courses, which provided specializations. For the Department of Relativity and Gravity, the program of special courses was compiled by Alexey Zinovievich himself in connection with the formation of a new department. \footnote{The second edition of the Program of Special Courses was prepared and published in 1983 by Yu. G. Ignatyev and S. P. Gavrilov.}. This was a very serious program, combining geometric and physical courses into a single whole.
\begin{enumerate}
\item Riemannian Geometry -- two semesters (once a week, exam) -- A.M. Anchikov;
\item Special Theory of Relativity -- one semester (once a week, test) -- V.I. Golikov;
\item Classical Theory of Gravity -- one semester (once a week, exam) -- K.A. Pyragas;
\item Lie Group Theory -- one semester (2 times a week, exam) -- R.F. Bilyalov;
\item General Theory of Relativity -- two semesters (2 times a week, exam) -- V.R. Kaigorodov;
\item Cosmology -- one semester (1 time per week, exam) -- G. Denisenko;
\item Experimental Justification of General Theory of Relativity -- one semester (1 time per week, test) -- K.A. Pyragas;
\item Special Seminar III -- V courses, weekly -- A.Z. Petrov;
\item Coursework, 4th year;
\item Industrial practice (differentiated assessment);
\item Diploma work.
\end{enumerate}

Petrov's department was very difficult in terms of studying, as was the adjacent department of theoretical physics. Therefore, AZ himself selected students for his department from a very large number of applicants, arranging for them, in fact, a test at the interview. If a student did not fit some parameters, AZ gently but persistently dissuaded him from this choice. Therefore, a lot of dreamy girls, despite their external data, were eliminated. As a result, a specialized group was formed from 15 students, only 3 of whom were female, but only one was a pretty blonde. It should be said that after the distribution in the third year, Petrov's son also appeared in our group (he did not study in our stream at the physics department), but he attended classes irregularly, and soon disappeared completely.

It is necessary to mention one more incident connected with assignment. Before assignment, Associate Professor I.S. Fishman called me for a talk to the Optics Department and tried in every way to persuade me to be assigned to this department, describing the future possibilities of such a choice. But I kept refusing. Then he asked, "What interests you in science?"\ I answered, "Plasma physics". Fishman immediately perked up and said that they were conducting large-scale research on plasma spectroscopy, and that this should suit me completely. In response, I said that I was not interested in the cold plasma that was being studied in the Optics Department, but in hot plasma, thermonuclear plasma, and relativistic astrophysical plasma. That was the end of our conversation. To the credit of Associate Professor I.S. Fishman, it must be said that neither the "dazibao"\ nor my refusal from the department affected my excellent grade in optics.
\subsection{Methods of Mathematical Physics and the Theory of Lie Groups}
I will tell another personal story, since it reveals some shades of Petrov's personality. This story lasted throughout the 4th and 5th semesters and is connected with the same teacher of the Department of Relativity and Gravitation Theory (let's call him N), but with different subjects that he taught. N was one of the students of AZ and a good mathematician, but at the same time he was characterized by a certain "stubbornness"\ and rude straightforwardness, due to which he was by no means a favorite among the students.

My conflict with N began during the summer session of the 4th semester, when I was taking an exam with him on the subject of "Methods of Mathematical Physics". N would let the whole group take the exam at once, as a result of which 25-30 people would be packed into a small classroom with tables in two rows. We would sit in a row of two people at a table with a common lower shelf. I knew mathematical physics well, so I was not at all worried, calmly writing down the answers to the questions on the exam ticket. Next to me on the left sat a girl P from our group, who was quite nervous and therefore kept a common notebook with lectures on mathematical physics on this very shelf, periodically modestly lowering her eyes and adding new details to her answer. Suddenly N unexpectedly began to walk around the class, looking into the lower shelves of the tables and even groping around with his hands. P got scared and with an elegant movement of her hand moved the notebook to me on the shelf. At that moment N came up to me and asked ominously, "What's this you've got there?"\ and pulled the lecture notes out from under the table. He quickly leafed through the notebook and, having made sure that they were his lecture notes, asked, "Is this your notebook?" "No," I said. "Then whose is it?"\ N asked angrily. "I don't know."\ "Then get out of my exam!" he shouted. I must say that I had never used cheat sheets, having been firmly brought up by my grandfather that cheat sheets were theft, and that in some places they chop off a hand for theft. Frankly, it was a shame. I had to retake the exam with bias, but N could not give me more than a B, branding me for life with the epithet "crook."

Then the 5th semester began and already a special course "Theory of Lie Groups", which to my "luck"\ was taught by the same N. The course was very difficult, and we solved the problems according to the book by N. G. Chebotarev "Theory of Lie Groups". These problems were original and the textbook was even, most likely, a monograph, presenting the personal research of the outstanding mathematician of Kazan University. It must be said that N knew the subject professionally, but he did not explain the solution to the problems, saying "Until you learn to solve all the problems from this book, you will not pass my exam."\ No one from our group solved these problems except me. I did it this way: I carefully copied all of N's lectures at home in special notebooks, put notes in the margins in those places where it was not clear, then rummaged through the books and gradually clarified these questions, writing the correct answers in the margins. In the end, I began to fully understand the question and solved the corresponding problem. I extended this method of taking lecture notes to other subjects, which helped me a lot in the future. When I came to the university, my classmates would copy my solutions into their notebooks. In January 1968, we took N on Lie group theory, and all my classmates thought that I would be the only one who would get an "excellent"\ grade. But things turned out quite differently. When I answered, N began to claim that I had copied everything. I asked him to question me, to give me a problem so that I could solve it in front of him. But N was adamant, did not ask any questions, and eventually gave me a "satisfactory"\ grade, accompanying this with the words -- "Well, so be it, I'll give you a C."\ I was completely shocked. This exam was the last one in the winter session, I had just gotten married and that day I had a plane ticket to Leningrad, where we were heading on our honeymoon. I lost my temper and left when I found out that all the students had gotten A's.

After returning from Leningrad, I approached N and asked to retake the exam, but he refused me in a harsh manner. I was shocked by the monstrous injustice and, as a last resort, went to Petrov. AZ was sitting alone at the desk at the department, smoking and writing something. I explained the situation to him, that I knew the subject well, "excellent", and considered the grade given to N unfair. AZ looked at me attentively with his deep dark gaze and asked if I was really sure that I knew the subject perfectly. After my affirmative answer, he said, "Well, sit down, I'll listen to you."\ He gave me a piece of paper and began asking questions quickly, easy ones at first, and then more tricky ones. "Well, you've confirmed your knowledge, let's have your record book."\ In the record book, he crossed out N's exam results in a sweeping manner and wrote "excellent"\ and his famous signature on a new line. Looking at me again kindly, he said, "Congratulations. Try to always defend the rightness of your convictions. This is a necessary quality of a scientist."\ My heart trembled with joy at the triumph of justice and was filled with some warm filial feelings for Alexei Zinovievich.

But even after this, N did not leave me alone, and all the years that I then worked at the Department of Relativity and Gravity Theory, he "pecked"\ at me wherever he could.
%

%
\subsection{Course and student scientific seminars}
An important component of the specialization in the Department of Relativity and Gravitation Theory were term papers and a student scientific seminar\footnote{It should be emphasized that the seminar had a scientific rather than an educational and methodological character.}. There were two term papers - one in the third year and one in the fourth. All students specializing in the department were required to attend a weekly student scientific seminar, which was led by Alexey Zinovievich himself. All teachers of the department also had to attend this seminar, which was quite impressive both in terms of numbers (about 50 people) and level. Thus, AZ created a close-knit educational and scientific team of the department and students, this talent of his is worth learning from. This AZ seminar should not be confused with the department's scientific seminar, which was held on Fridays and therefore was called "BF"\ - Black Friday, the second half of the name was caused by a very difficult situation for the speaker, whose report was subjected to a thorough and unflattering analysis.

\FIG{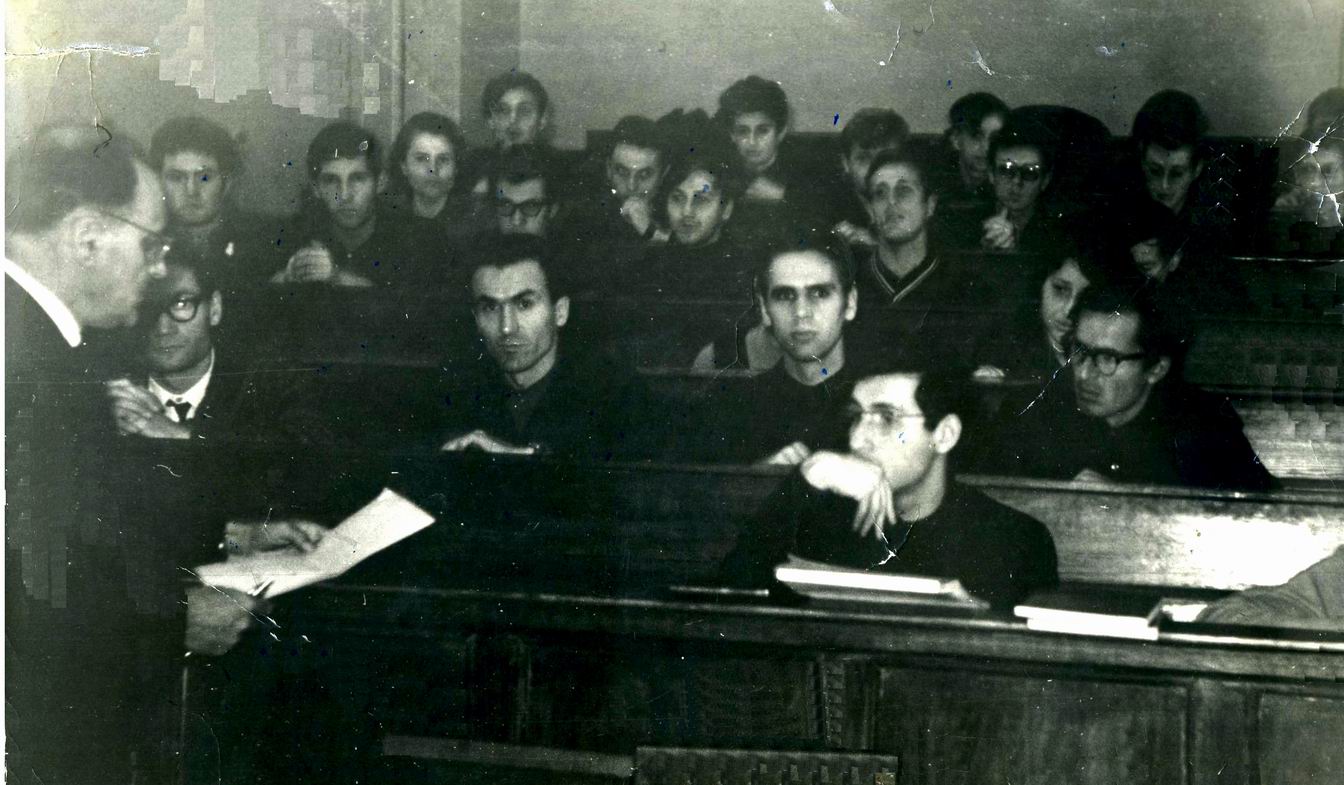}{14}{Student seminar under the direction of A.Z. Petrov (in the foreground). 1968, Physics Auditorium No. 3. In the first row is a fifth-year student, presumably the speaker. In the second row, from left to right: Associate Professor V.I. Golikov, Assoc. Professor A.M. Anchikov, Assoc. R.S. Singatullin, Assoc. S.L. Tsarevsky. In the fourth row are five of my third-year classmates, with I. Mubarakshin on the left. In the fifth row, second from the left is Yu. Ignat$'$ev. In the sixth row, first from the left is A. Pestov.}

It should be noted that, apparently, the term paper in the third year was a "test run"\ and was done outside the program. All term papers and theses were presented at a student scientific seminar; also, during the preparation of the theses, AZ assigned students to write review reports on the research topic. This assignment had to be taken very seriously, raising a significant number of foreign publications. For example, when doing my thesis, I had to translate from English about 20 articles from Phys. Rev., Nature, and other publications. I carefully copied the translations into notebooks and filed them in a common folder. I will note that all the main international scientific journals republished in the USSR, which did not mutually recognize copyright until 1975 \footnote{Thanks to this, all the most significant foreign scientific journals and monographs were freely available to students and researchers.}, were located right in the reading room on the second floor of the main university building. If the publications were not available, we ordered copies of them via the IBA (interlibrary loan). In this way, we were taught to work seriously with scientific literature. In addition, AZ had an extensive personal scientific library at the department\footnote{which he later took with him to Kyiv}, consisting mainly of reprints of articles, preprints and dissertations that were sent to him from all over the world by Authors, accompanying them with dedications. We also often used this literature, leaving inserts with a receipt with the department lab assistant.

Giving a report at a student scientific seminar was quite difficult, it required very serious preparation, considering the deep and insightful mind of AZ, his extensive knowledge and strict, although delicate attitude towards young speakers. AZ never tried to humiliate the speaker, to show his true place, as some of our candidates for doctorate, lovers of "cutting"\ the speaker, like to do, which I often observed at some seminars in the capital. AZ, of course, treated the speaker with respect, putting the process of moving towards establishing the truth in the first place.

\subsection{Kinetic equation}
In my fourth year, I did a review of general relativistic kinetic theory based on the works of Nikolai Aleksandrovich Chernikov and at the same time tried to generalize the general relativistic kinetic equation to the case of additional electromagnetic interaction. The kinetic equation obtained as a result of the generalization had the form
\begin{equation}\label{eq1}
p^i\frac{\partial}{\partial x^i}f(x,p)+\left(\Gamma^\alpha_{ik}p^ip^k-\frac{e}{m}F^\alpha_{.~k}p^k\right)\frac{\partial}{\partial p^\alpha}f(x,p)=I(x,p),\quad (\alpha=\overline{1,3},\ i,k=\overline{1,4}).
\end{equation}
where $f(x,p)$ is the invariant distribution function with respect to the coordinates of the seven-dimensional phase space $\{x,p\}$, $\Gamma^\alpha_{ik}$ are the Christoffel symbols of the second kind, $F^\alpha_{.~k}$ is the electromagnetic field tensor, $I(x,p)$ is the invariant collision integral.

After I wrote the equation \eqref{eq1} on the board, Alexey Zinovievich immediately asked me -- "Why is this equation written in a non-invariant form? After all, the distribution function is invariant, and there is an invariant on the right-hand side of the equation". I could not answer this question. My course supervisor, Z, himself an ungraduated teacher, also could not answer this question. AZ made a polite remark to the manager, and told me, "Get busy proving the invariance of this equation, or present its correct version."\ At this point, the hearing of my report was terminated.

It should be said that the preprints of N.A. Chernikov's articles, as well as his doctoral dissertation "Kinetic Theory of Relativistic Gas", were taken from Petrov's library by my supervisor. As I understood much later, AZ was fully aware of Chernikov's works, since, firstly, they were friends, and, secondly, Chernikov presented his doctoral dissertation in 1963 at the Department of Relativity and Gravitation Theory. I think that AZ asked Chernikov a similar question, but, judging by the question to my report, he also did not receive a satisfactory answer from him. And although the answer to the question lay almost on the surface and, if the question had been properly formulated, AZ, as a Kazan geometer, could have answered it himself, apparently, the insufficiently clear formulation of the problem did not allow him to do so then.

The question that had stumped me forced me to seriously study the solution of this problem. I spent almost the entire spring semester of 1969 struggling with it. My reasoning went something like this: since the distribution function is invariant, and $p^i$ is a vector, then $f(x,p)$ can depend on this vector only through convolutions of the form $A_n\equiv a_{i_1,\ldots,i_n} p^{i_1}\cdots p^{i_n}$, where $a_{i_1,\ldots,i_n}$ is the covariant tensor of valence $n$ of the Riemannian space -- $f=f(A_0,A_1\ldots,A_n)$. But the action of the kinetic operator (the left-hand side of equation \eqref{eq1}) on $A_n$ is a scalar. Therefore, the entire equation \eqref{eq1} is invariant when acting on the invariant distribution function. I reported these considerations at a student seminar in April 1969, for which I was awarded stingy praise from Alexey Zinovievich and an "excellent"\ grade for my coursework.

I will note that in fact, the problem was solved more simply - for this it was necessary simply to understand that the phase space is a vector bundle, and the left-hand side of the equation \eqref{eq1} without the electromagnetic term is the Cartan derivative in this bundle. It is strange that Chernikov himself did not use this fact, although he claimed in his dissertation that the phase space is a vector bundle. AZ himself, of course, simply did not notice this fact, perhaps because of the cumbersomeness of the notation, although he undoubtedly knew about it, being one of the leaders of the Kazan geometric school and knowing first-hand the works of E. Cartan and B.L. Laptev. Or, perhaps, he wanted me to come to this conclusion on my own. One way or another, I finally came across the excellent book by A.A. Vlasov "Statistical Distribution Functions, 1966", which brought final clarity to this issue.

The time for vacation was approaching, and at the same time, the distribution of topics for diploma theses. From the topics proposed at the seminar, I chose the work under the supervision of AZ, "Immersion of the gravitational field ... in Euclidean space."\ This topic was the only one proposed by AZ and, frankly speaking, I did not like it very much, since it boiled down to solving a large number of constraint equations, which was very cumbersome and of little use to me. But I really wanted to work with Petrov and stay with him in graduate school. At the beginning of May, AZ called me in and said that due to newly arisen circumstances, he would not be able to supervise the diploma thesis in the next academic year and suggested that I go to N.A. Chernikov in Dubna to continue this work under his supervision. "They will not be able to help you in this direction here. We do not have such specialists. And two of our graduates are already working at JINR -- Erik Tagirov and Aivengo Pestov."\ -- Alexey Zinovievich explained and gave me a letter of recommendation, wishing me success and adding that I could always count on his support in the future.

\subsection{Visit to Dubna: 1969}
So, at the end of May, completely light, armed with a letter of recommendation, a student card and a scholarship I had just received, I left for Dubna. First to Moscow, and then from Savyolovsky Station to Dubna.
I must say that in those years I was very easy-going, frivolous and had no idea about any formalities of settling in. At one time, like many of my peers, I watched the film by director Mikhail Romm "Nine Days of One Year"\footnote {The main action of which, as I understood upon arrival, took place in Dubna, JINR.} we all wanted to become nuclear physicists and stand chest-deep in a stream of neutrons - such were the times then - times of general enthusiasm and thirst for discoveries. Dubna, like nothing else, helped me in the implementation of these plans. Inspired by the familiar sight of the cafe "Under the Integral" from literature, I briskly walked through Dubna to the coveted gates of the Joint Institute for Nuclear Research (JINR), where an armed guard at the checkpoint cooled my ardor, explaining that a student card was clearly not enough for me to enter the territory of the closed scientific center, I needed an application, access, a passport and all that. But even if I had a passport with me, the paperwork would have taken a couple of days, and I had a return ticket "Moscow-Kazan"\ for the next day. \footnote{In those "terrible"\ years, a passport was not required for travel on the train.}.

But as I said, I was frivolous, but also optimistic and daring. No security guard could stop me. I headed along the high, serious fence surrounding JINR. Chains jingled and dogs barked behind the fence. But I was stubborn. Finally, I discovered an inconspicuous path in the thickets, running across the railroad tracks right to the fence - apparently, local employees, not wanting to make a huge circle on the way to the checkpoint, trampled it. The path was supposed to lead to some entrance. And so it was. Having moved the fence boards aside, I safely climbed onto the territory of the closed, albeit united, institute. Here I wandered for quite a long time between the buildings, trying to find the building of the LTP (Laboratory of Theoretical Physics). About an hour and a half later I found it. Inside, too, there were many rooms, but the noise of a lively conversation came from only one room. A group of middle-aged and older people were sitting at a table during a friendly scientific meal. On the wall hung a blackboard covered with chalk formulas and it was very smoky. I timidly asked where I could find Professor Nikolai Alexandrovich Chernikov. In response, a rather heavy and tall man rose from the table and asked menacingly, "It's me, what do you want?"\ I explained. He called somewhere on the phone, and soon Ivanhoe Pestov arrived. Chernikov told him, "Here's a student from Alexei Zinovievich. Go have a bite to eat somewhere, and after dinner take him to my home."

In a two-story cottage among the pines, I was received by Chernikov's wife, a beautiful and friendly woman. She treated me to tea and said, "I'll make a bed for you on the second floor. Nikolai Alexandrovich asked you to wait until the morning, since he will be coming home late from work."\ There was nowhere to go, I walked around Dubna a little, reached the famous cafe "Under the Integral" on the station square, fortified myself there, including ice cream and a cup of coffee, and then went "home". Chernikov's wife sent me to bed, comforting me with the words, "Talk to me in the morning."\ In the morning, Nikolai Alexandrovich was not entirely friendly. After listening to me again and reading the letter from Petrov, he sullenly said, "You have one way to start working with me. This is to pass exams at Moscow State University and Joint Institute for Nuclear Research in the subjects: Physics of the Atomic Nucleus, Quantum Field Theory, Plasma Physics, Physics of Elementary Particles, Quantum Statistics. Study in the summer and come and take them."\ I was dumbfounded -- "Five fundamental and completely unfamiliar subjects in the summer?!"\ -- "Well, then, come back next year!"\ -- Chernikov retorted. That was the end of my audience. I returned to Kazan in a depressed state -- the shining, as it seemed to me then, peaks of science were covered in clouds of fog, suddenly becoming unfriendly and completely inaccessible.

\subsection{Double Degree: 1969 -- 1970}
In September I met with Alexey Zinovievich and explained the situation to him. He frowned in bewilderment and said, "Yes, it is rather difficult to master such a layer on your own."\  I already knew that AZ was leaving us for Kyiv and did not know where to put my feet. I asked him what I should do. In response, he announced his decision to move to Kyiv and said that the department did not have any postgraduate studies planned for the next year, and there were big problems with its management. He advised me to contact Professor Uno Hermanovich Kopvillem at the Physics and Technology Institute -- "He is a real theoretical physicist and works, among other things, on issues of detecting gravitational waves. I think this is what you need."\ I timidly asked AZ if there was any possibility for me to go with him to Kyiv. To this he replied that his own position at the ITP was not yet completely clear, so he could not guarantee anything yet. But let's write to each other and I'll try to do something. Write to me in Kyiv. He wrote his Kiev address on a piece of paper, said goodbye to me warmly, and, probably noticing my extremely pitiful appearance, called Kopvillem on the phone, asking for me.

That's how I met Uno Hermanovich, a remarkable man and scientist, a friend of AZ. UH, as Kopvillem was called at the Phystech, was a very extraordinary person, cheerful, athletic and not without a certain eccentricity. He promised to take me on as a graduate student, but at the Aviation Institute, where he worked as a professor at the department, I think, of electromagnetism, for 0.5 of a position. It turned out that the head of this department asked to place his son in the Phystech graduate program (UH was a very famous scientist), so Kopvillem's postgraduate place was taken. Well, I decided to go for this adventurous option. Uno Hermanovich set me a task from the field of theoretical physics, very far from the theory of gravity - it was the theory of double gamma-gamma correlations in the decay of nuclei, gave me the basic literature and assigned me a lot of consultations. I mastered this topic (and it turned out, like any positive knowledge, to be useful in the future), even making a report at a conference together with UH, and wrote a diploma thesis, but did not defend it. At the same time UH suggested to me: "Let's finally quantize the gravitational field"\ and gave me a reprint of Gupta's famous work. But these plans were not destined to come true... However, Uno Hermanovich and I maintained close friendly relations throughout our lives.

At the same time, not wanting to block my path to the theory of gravity, I wrote my diploma thesis with AZ's closest student (as Alexey Zinovievich introduced him to me) -- associate professor Vladimir Vladimirovich Kaigorodov (VR) on a topic that was simultaneously close to the gravitational experiment and to the theory of modeling the gravitational field -- "The Electromagnetic Character of the Gravitational Field". Petrov himself set the task and passed it on to Kaigorodov -- to reveal, on the basis of Einstein's equations and gedesics, in a weak non-relativistic approximation, the common features of gravitational and electromagnetic interactions.

VR outlined the task for me in general terms and told me what days and times he could be found at the department for consultations. I would like to note that it was quite difficult to get a normal consultation at the department due to the small area of ??the department premises and the large number of teachers and students. Therefore, after trying a couple of times, I gave up this activity and began to work independently, as I was accustomed to. Usually, I sat in the reading room, where I read articles and worked. By the end of the calendar year, the diploma thesis was ready in rough form, all its results were obtained. I managed to reduce the entire system of Einstein equations and geodesics in a weakly relativistic approximation to almost the equations of electrodynamics, in which some additional currents and dissipative forces appeared. At the end of December, I showed up at the department, since I needed to get a grade for industrial practice. "Where have you been!"\ -- my supervisor cried out. "We were about to submit you for expulsion!"\  However, after seeing the draft version of my thesis in my hands and reviewing its results, the VR softened, gave me an "excellent"\ mark for industrial practice and said, "But you should have warned me!"\ In May, I defended my thesis with an "excellent"\ mark. The reviewer was Associate Professor Kazimir Antonovich Pyragas.

\subsection{Departure of AZ and correspondence: 1970 -- 1973}
As far as I can tell, Alexey Zinovievich finally moved to Kyiv in the summer of 1970. I haven't seen him since. At the end of the summer of 1970, I had some extraordinary family circumstances and a complete "reset"\ of my family life, which initially pushed the issues of my scientific career into the background. In connection with this, at the end of September, I left first for Dnepropetrovsk, then for Kharkov, and finally for my small homeland - Lvov. My grandmother and aunt lived there in a three-room apartment, and I planned to register there with my wife in order to be closer to Kyiv. Before leaving, I wrote to AZ, introducing him to my plans. In response, he wrote that there was a scientific group at the Ivan Franko University in Lvov that was working on issues close to the theory of gravity, and he advised me to contact them. He also suggested that I regularly come to Kyiv for consultations at the ITP (Institute of Theoretical Physics) during the year, so that I could enroll in graduate school the following year. However, this "brilliant plan"\ of mine failed miserably. In those years, Lvov had the status of a border city, and therefore registration was only possible with a wife, but my wife was already working at a very closed enterprise in Dnepropetrovsk, as a result of which the registration failed.

I had to return to Kazan, where admission to postgraduate studies had already ended, and so I had to ask Professor (then Associate Professor) Shamil Shagivaleevich Bashkirov (for whom I had worked on a voluntary basis during my student years, performing some radiation measurements and calculations) to take me to his Department of Solid State Physics. Thus, from December 1, 1970, I became an engineer at the Nuclear Physics Laboratory of Kazan University. At the same time, the very difficult process of detaching a young specialist (my wife) from a restricted enterprise in the USSR began.

Finally, in December 1970, I finally came to my senses and wrote to Aleksey Zinovievich about my ordeals and asked for advice on what to do in this situation if I wanted to modelling study gravity theory while continuing my research in the field of general relativistic plasma theory. AZ quickly replied that it would be most convenient to contact A.P. Shirokov, who at that time became the head of the Department of Relativity and Gravitation (TRG) at Kazan University. At the same time, AZ wrote to Shirokov asking him to assign me to the department and provide scientific supervision. After that, I came to the TRG department and talked to A.P. Shirokov. Aleksandr Petrovich, as usual in such cases, blushed and began to explain that he was a geometer and simply could not supervise research in such a direction. I explained to him that I did not need scientific supervision, I would carry out scientific research independently, but I asked him to at least formally become my scientific supervisor and give me the opportunity to present reports at the department's seminars\footnote{The General Scientific Student Seminar ceased to exist with the departure of AZ and never resumed its work.} In response, Shirokov blushed again and refused in a soft but unconditional manner: "How can I be a formal scientific supervisor in a field of science in which I am not a specialist?"\  But he did allow me to attend the department's scientific seminar.

I wrote to Alexey Zinovievich about the current situation, to which he philosophically replied that I should not give up, I should work, and get the necessary advice from Uno Hermanovich Kopvillem, and from him himself, if necessary. I did so, periodically sending AZ letters with my results and questions. AZ always responded very quickly, in small, clear handwriting on notebook pages, and these laconic answers of his helped me a lot. It should be said that it was he who inspired me to think about the derivative in stratification. In addition, I wrote him a lot of personal things, since I was young, inexperienced and hot-blooded, having lost my father in 1968, experiencing acute material deprivation and needing a mentor. In a sense, Alexey Zinovievich replaced my father until 1972, sometimes with advice, sometimes with lively sympathy, sometimes with strict instructions. How can I forget his deep, serious, expressive eyes with permanent dark circles under them, his kindness and exactingness, clarity and depth of thinking, fatherly warmth? But this is already too personal. He helped me a lot with his letters in those difficult years for me. In my youth, I did not understand then that I was burdening this very busy and sick man, a world-class scientist and an outstanding leader, with my stupid youthful letters.

\section{The scale of Petrov's personality as a scientist and a person}
At the end of the article I will try to assess the scale of the personality of Alexey Zinovievich Petrov. Without establishing this assessment, a true history of science is impossible. There are different assessments of the scale of AZ's personality, many of which are caused by ambition and professional jealousy, clan interests of various groups of scientists, as well as the well-known saying "A prophet is not without honor except in his own country and in his own house"\ (Gospel of Matthew 13:53). In general, the high position of A.Z. Petrov has been firmly established in world science as a geometer who made a huge contribution to the development of mathematical methods for studying gravitational fields, in particular, who created an algebraic classification of gravitational fields (the famous "three Petrov types and one degenerate") and a classification of gravitational fields by groups of motions. Each fundamental foreign, as well as Russian, textbook (or monograph) on the general theory of relativity contains the corresponding sections. It is enough to name, for example, the widely known to specialists monograph by D. Kramer, H. Stephany, M. Maccallum, E. Herlt edited by E. Shmutzer ``Exact Solution of the Einsteins Field equations'', Deutscher Verlag der Wissenshaften, Berlin, 1980, the famous textbook by L.D. Landau and E.M. Lifshitz "Field Theory"\ and many others. Without these results of Petrov it is impossible to imagine the modern theory of gravitation. This is evidenced, for example, by the fact that the most popular applied mathematical package among theorists at present ``Maple'' contains special software procedures, such as, for example, \texttt{PetrovType} for calculating the type of gravitational field according to Petrov. These facts alone allow us to give a cautious assessment of Alexey Zinovievich Petrov as an outstanding scientist, geometer and theoretical physicist of the second half of the 20th century.

Other formal facts confirming this assessment are the following: he was awarded the Lenin Prize in 1972 \footnote {And the Lenin Prize was the highest State Prize of the USSR, which was almost twice as large as the modern Russian Federation and much larger in scientific and technical potential. It is enough to point out the following fact: gravity conferences in the USSR brought together about a thousand researchers, while similar conferences in the Russian Federation brought together less than a hundred.} for the invariant-group methods for studying gravitational fields that he developed, and his election as an academician of the Academy of Sciences of the Ukrainian SSR in 1969. In addition, it is necessary to note A.Z. as an outstanding organizer and leader of scientific research. This is evidenced by the following facts: the chairmanship of A.Z. Petrov in the Soviet Gravitational Commission throughout its existence, the creation (in 1960) and development (1960-1970) of the only department of relativity and gravitation theory in the Soviet Union, the leadership of the department of relativity and gravitation theory of the Institute of Theoretical Physics of the Academy of Sciences of the Ukrainian SSR and the organization (together with Braginsky) of the experimental search for gravitational waves. In addition, one can also assess the scale of A.Z. Petrov's personality as outstanding, given his incredibly difficult path from a peasant boy to the heights of science through orphanhood, deprivation, malnutrition, injury in the Great Patriotic War, lack of normal housing conditions (\cite{Amin}) and the constant exhausting intrigues of colleagues (\cite{Vladimir_KazCas}, \cite{Vladimir_STFI}).

\subsection{Different opinions}
One of the direct consequences of the saying "On prophets in the fatherland..."\ is the insufficiently high assessment of A.Z. Petrov as a scientist-geometer in the environment of Kazan geometers, where the opinion about the genius P.A. Shirokov has taken root, who suggested to the student A.Z. Petrov the topic of the diploma work "Einstein Spaces", which allegedly grew into his two monographs. On the other hand, there is also an opinion here that in his classification of Einstein spaces he simply generalized the already known results on spaces with a certain signature of the metric. They say he was simply lucky with the necessary application to physics. In the materials placed on the stands of the Faculty of Mathematics and Mechanics named after N.I. Lobachevsky of Kazan University, the name of A.Z. Petrov is personified only as a student of P.A. Shirokov and a laureate of the Lenin Prize, received for solving a physical problem. For 8 years I worked as the head of the department at this faculty, and I also "sat"\ in the doctoral council for geometry since 1995, so I managed to feel the spirit of the somewhat dismissive attitude of Kazan geometers to the personality of Petrov during these 23 years. I felt it especially vividly when lecture hall No. 610 of the faculty began to be equipped as an assembly hall and portraits of all the outstanding mathematicians of Kazan University were hung in it. Of these approximately 10 mathematicians, only one N. G. Chetayev was a laureate of the Lenin Prize and N. G. Chebotarev was posthumously awarded the Stalin Prize. At the same time, none of them was an academician. The other mathematicians represented in the portraits were authorities of the "local level", i.e., at best, famous scientists. The portrait of Alexey Zinovievich Petrov was not among them. I went to the director of the institute and asked why the portrait of the most famous mathematician of the 20th century at Kazan University, A.Z. Petrov, was not on the walls of the auditorium. In response, the director explained that we had consulted with the management of all the mathematical departments in the directorate and had finally approved this list. I replied that this was unfair and that I would insist on attaching A.Z.'s portrait to this auditorium. Since the director resisted, I asked my student, now the head of the geometry department, A.A. Popov, as well as the head of the TOG department, to support my request. As a result, after lengthy negotiations, the portrait of Alexey Zinovievich was finally hung in auditorium 610.

Here is a living <<statistical>> example of this attitude to AZ among geometers - in a solid article in volume, dedicated to the life and work of A.Z. Petrov and the 100th anniversary of Petrov, \cite{Amin_2} (see also \cite{Amin}) the adjective <<brilliant>> is used 1 time (but only in relation to P.A. Shirokov), 4 times "talented"\ (but in relation to the students of Chebotarev, Petrov and the young participants of the Petrovsky Readings), 1 time "outstanding"\ (about the lecturers of the Petrovsky Readings), 1 time "capable"\ (this is about Petrov from the lips of associate professor V.G. Kopp), 8 times brilliant (6 times about Petrov's teachers, 1 time about Petrov the schoolboy and 1 time about Petrov the lecturer). The characteristic "fame"\ is used about Alexey Zinovievich twice, but "persistent"\ is used 6 times, "persistent"\ is used 2 times, and "purposeful"\ is used 1 time, and also "will"\ is used. As someone said, "facts are stubborn things!"

A beautiful picture is emerging? The poor, untalented boy from a peasant family was very lucky - he found himself in the company of brilliant and talented teachers and equally talented students. He was given a very promising task. Thanks to his will, tenacity, persistence and determination, he was still able to generalize the previously known task and, by a happy coincidence, became an academician and a Lenin Prize laureate! Just like some famous fairy tale by Hoffmann about the Nutcracker! This is his vision of the world due to the reasons listed at the beginning of the section.
\subsection{A.Z. Petrov: from geometry to physics}
What is the uniqueness of the personality of Alexey Zinovievich Petrov, a scientist and a person? The scientist Petrov stubbornly went all his life to comprehend the secrets of the universe, he was interested in how the real world is structured, and not in solving elegant mathematical puzzles. And on the way to comprehending the truth - the secret of this world, he used all his talents - the talents of the mind and soul, the talents of a mathematician, a leader and an organizer. On this path, he did not spare either himself or his enormous life energy and burned out like a bright star, flashing in our sky. We can say that AZ repeated the life trajectory of his brilliant predecessor - Nikolay Ivanovich Lobachevsky, visibly once again confirming the correctness of the axiom of the English playwright "Life is a tragedy". Lobachevsky also started from geometry, but at the same time his driving force, the motivation of his research was always the geometry of the real world. There is a lot of evidence of this, scattered in the form of statements in various treatises of Lobachevsky. He posed the question of the connection between mechanics and geometry and expressed an assumption about the possible connection between mechanical forces and distances. He left another visible evidence to Kazan University in the form of an astronomical observatory built according to his drawings, which still adorns the territory of Kazan University. Reading through the works of Lobachevsky, which, among other things, describe the scheme of a possible experiment on stellar triangulation of what he modestly called the "imaginary world", you begin to understand the deep meaning of the observatory he built. And he was on the right track, but at that time he did not realize the full technical complexity of such an experiment, carried out already in the 21st century. Petrov also tried to build his own "gravitational wave observatory"\ first at Kazan University and then at the Institute of Theoretical Physics in Kyiv. And just like Lobachevsky, Petrov did not realize the technical complexity of the experiment, which was also carried out in the 21st century. Just like Lobachevsky, A.Z. Petrov was subjected to unspoken obstruction after his death, precisely by his fellow geometers. As they say, "the mortirologies of the holy martyrs are as alike as two peas in a pod!"\ This movement toward the Truth alone distinguishes both Lobachevsky and Petrov from the vast mass of mathematicians and makes them outstanding mathematicians. Lobachevsky is also worthy of the epithet "brilliant"\ for the breakthrough he made in the scientific worldview, and Petrov is "outstanding"\ for his discovery of new patterns and the methods he developed for studying an existing theory. The same path was paved by other outstanding mathematicians who posed mathematical problems based on real-world problems: Helmholtz, Euler, Newton, Lagrange, Hamilton, Hilbert, Weyl, Noether, Lyapunov, Kolmogorov, Pontryagin, Arnold and many others. Only the infinitely multifaceted world around us is capable of deeply and truly puzzling scientists and directing their efforts to solving great mathematical problems.

\FIG{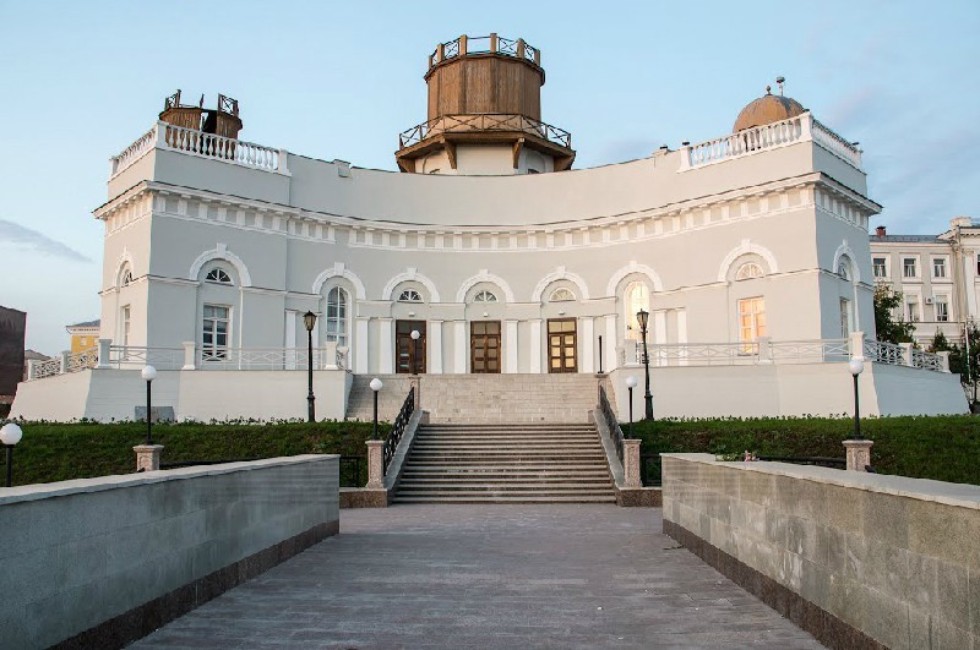}{14}{The Kazan University observatory, where Lobachevsky attempted to measure the sum of the angles of stellar triangles.}

\subsection{From an interview with Yu. G. Ignat$'$ev for the magazine "Matrix"}
-- In my opinion, science is a type of conscious activity for studying the World in the broadest sense of the last word. At the same time, the criterion of science should always be an experiment, in which the adequacy of a mathematical model is tested, and the mathematical model itself should predict the future state of the object under study, based on knowledge of its current state. Along with science in this understanding of mine, pseudoscience also actively exists, the forms of which are very diverse, but which do not satisfy precisely the specified criterion. There are so-called ideological (philosophical) sciences, in which there is no criterion of truth, except for the criterion of logic in accepting a given system of axioms (views). A special place here is occupied by mathematical sciences, in fact, with the same criteria of truth as ideological sciences. All these sciences were once generated as a result of the process of man's cognition of the World, but later they received their internal development. I may express a seditious thought, which I think will not be very much liked by "pure mathematicians"\ who defend their right to develop their mathematical fields in any direction, without regard to the demand for such research and justifying their activities by the need for internal development of theory. Mathematics is often called the "Queen of Sciences". To some extent, this is true, because without mathematical justification it is impossible to build any adequate model of the World, which would make it possible to predict its properties and compare "figures"\ with reality. But I think that in such cases the term "Servant"\ is more acceptable than "Queen". The tool/set of tools for building models and research is one of the most adequate languages of communication between people engaged in scientific activity. The development of language is a noble task. Indeed, if it is strictly proven that under certain conditions the logical chain A$\Rightarrow$B$\Rightarrow$C$\Rightarrow\cdots$Z is valid, then the researcher under these conditions immediately uses the shorter logical chain A$\Rightarrow$Z. This is the development of language. But the trouble is that many pure mathematicians follow the path of least resistance in their research: instead of studying a certain chain that interests the researchers, they either add new words to their language, or new parts of speech, or new rules of grammar and continue their game of glass beads.
\section*{Conclusions}
To sum up this article, we will highlight its main position: \\[12pt]
Alexey Zinovievich Petrov is an outstanding geometer and theoretical physicist of the second half of the 20th century, one of a small number of stars of the Soviet Union's gravitational theorists, and certainly the second brightest (after Lobachevsky) star of the Kazan geometric school. I am sure that time will confirm this assessment.

\end{document}